\documentclass{article}
\usepackage{spconf,amsmath,graphicx}
\usepackage{subfigure}
\usepackage{adjustbox}
\usepackage{booktabs}
\usepackage{mathtools}
\usepackage{amssymb}
\usepackage{amsmath}
\usepackage{multirow}
\usepackage{makecell}
\usepackage{wrapfig}
\usepackage{hyperref}
\usepackage{xurl}
\usepackage{cleveref}

\title{CrossSpeech: Speaker-Independent acoustic representation \\for Cross-lingual Speech Synthesis}
\name{Ji-Hoon Kim, Hong-Sun Yang, Yoon-Cheol Ju, Il-Hwan Kim, Byeong-Yeol Kim} 
\address{42dot Inc., Seoul, Republic of Korea}

\begin{document}
\ninept
\maketitle
\begin{abstract}
While recent text-to-speech (TTS) systems have made remarkable strides toward human-level quality, the performance of cross-lingual TTS lags behind that of intra-lingual TTS. This gap is mainly rooted from the speaker-language entanglement problem in cross-lingual TTS. In this paper, we propose CrossSpeech which improves the quality of cross-lingual speech by effectively disentangling speaker and language information in the level of acoustic feature space. Specifically, CrossSpeech decomposes the speech generation pipeline into the speaker-independent generator (SIG) and speaker-dependent generator (SDG). The SIG produces the speaker-independent acoustic representation which is not biased to specific speaker distributions. On the other hand, the SDG models speaker-dependent speech variation that characterizes speaker attributes. By handling each information separately, CrossSpeech can obtain disentangled speaker and language representations. From the experiments, we verify that CrossSpeech achieves significant improvements in cross-lingual TTS, especially in terms of speaker similarity to the target speaker.

\end{abstract}

\begin{keywords}
text-to-speech, speech synthesis, cross-lingual TTS, speaker generalization
\end{keywords}

\section{Introduction}
\label{sec:intro}
Due to the development of deep learning, text-to-speech (TTS) systems have made dramatic progress towards human-level quality~\cite{shen2018natural,lancucki2021fastpitch,kim2021conditional,tan2022naturalspeech,ju22_interspeech}. However, the progress is only applicable to intra-lingual TTS. The quality of cross-lingual TTS, which aims to generate natural speech in one target language from a source speaker using another language, still lags behind the quality of intra-lingual TTS. This degradation in cross-lingual speech is mainly rooted from speaker-language entanglement. In practice, it is common for one source speaker in a training dataset to speak one source language, and the speaker identity is prone to stick to linguistic information. Therefore, it is hard to preserve the speaker identity when the source language representation is replaced with the target language representation.

A number of works have been made to improve the performance of cross-lingual TTS by mitigating the speaker-language entanglement problem. They can be divided into two broad categories~\cite{xin2020cross,zhan21_interspeech}. One is to utilize language-agnostic text representation which can be shared across multiple languages. For example, B. Li \textit{et al.}~\cite{li2019bytes} introduces ``byte" representations for English, Mandarin, and Spanish. H. Zhan \textit{et al.}~\cite{zhan21_interspeech} utilizes International Phonetic Alphabet (IPA) and verifies the superiority of using IPA over language-dependent phonemes in cross-lingual TTS. Extending IPA, M. Staib \textit{et al.}~\cite{staib2020phonological} and F. Lux \& N.-T. Vu~\cite{lux2022language} present articulatory features which enable the model to maintain its topology for different languages. Another approach presents training techniques to learn disentangled speaker and language information. Y. Zhang \textit{et al.}~\cite{zhang2019learning} leverages domain adversarial training to remove speaker information from text encoding. D. Xin \textit{et al.}~\cite{xin2021disentangled} adds mutual information minimization loss to prevent speaker embedding from leaking into language embedding. Recently, SANE-TTS~\cite{cho2022sane} applies speaker regularization loss on non-autoregressive TTS model~\cite{kim2021conditional} to avoid speaker bias of text embedding in duration predictor. 

Although previous studies have tried to improve the quality of cross-lingual TTS by decomposing the speaker and language information, the decomposition is limited to the level of input tokens. Even though the speaker and text representations are separated in the input token space, they are combined at the input level of ``decoder" to produce acoustic representations, which leads to speaker-language entanglement again. In this paper, we propose CrossSpeech which highly improves the quality of cross-lingual TTS by disentangling speaker and language information in the level of decoder output frames. To this end, we decompose the speech generation process into the speaker-independent generator (SIG) and speaker-dependent generator (SDG). From the text inputs, SIG produces the speaker-independent acoustic representation through the mix-dynamic speaker layer normalization, speaker generalization loss, and speaker-independent pitch predictor. On the other hand, SDG models the speaker-dependent acoustic representation via the dynamic speaker layer normalization and speaker-dependent pitch predictor. Experiments demonstrate that CrossSpeech highly improves the quality of cross-lingual TTS, particularly in terms of speaker similarity to the target speaker.

\section{model architecture}
\label{method}

\begin{figure*}
    \centering
    \subfigure[Overall architecture]{
        \includegraphics[width=0.575\columnwidth]{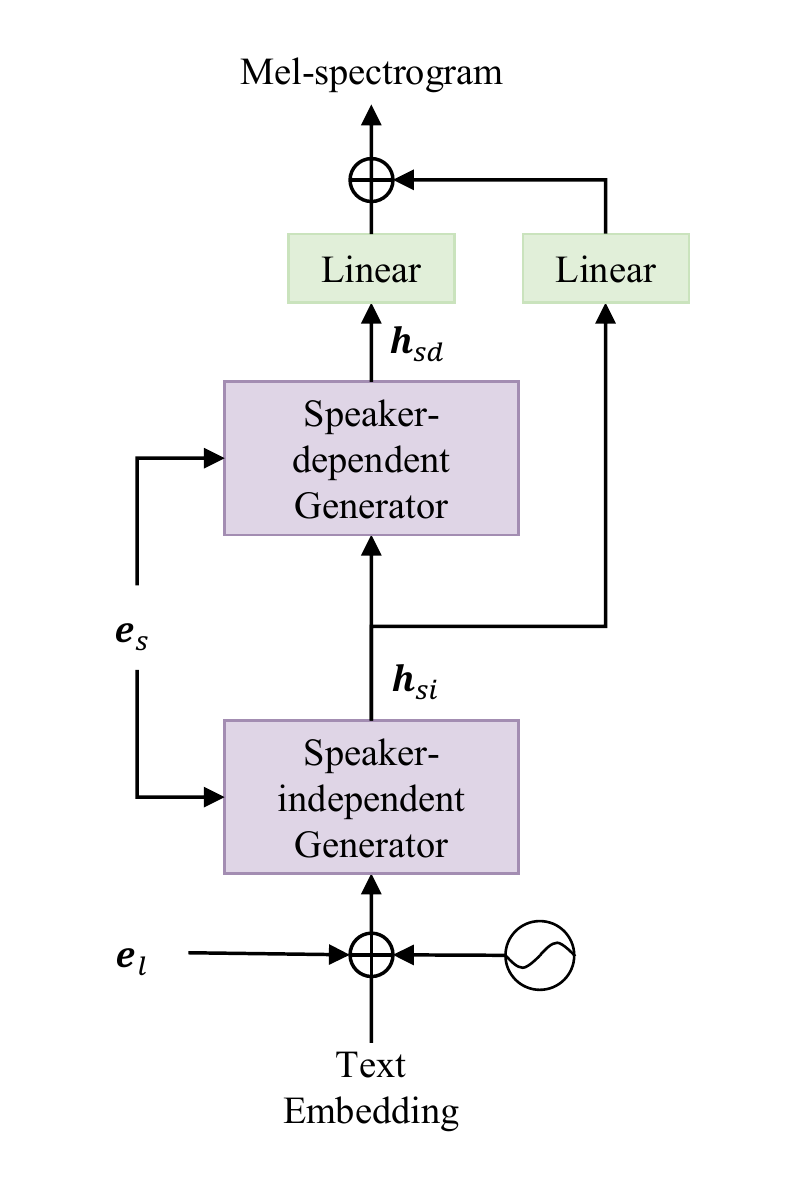}
        \label{fig:overall}}
    \subfigure[Speaker-independent Generator]{
        \includegraphics[width=0.575\columnwidth]{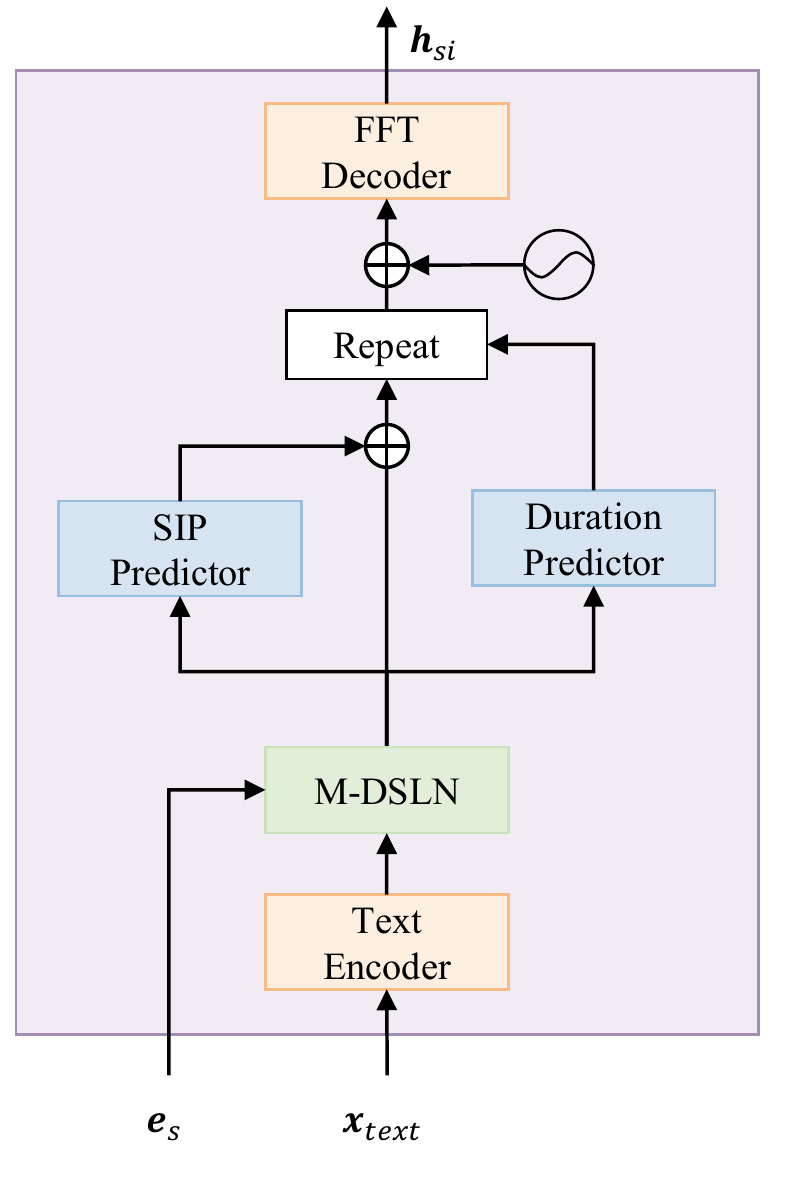}
        \label{fig:sag}}
    \subfigure[Speaker-dependent Generator]{
        \includegraphics[width=0.575\columnwidth]{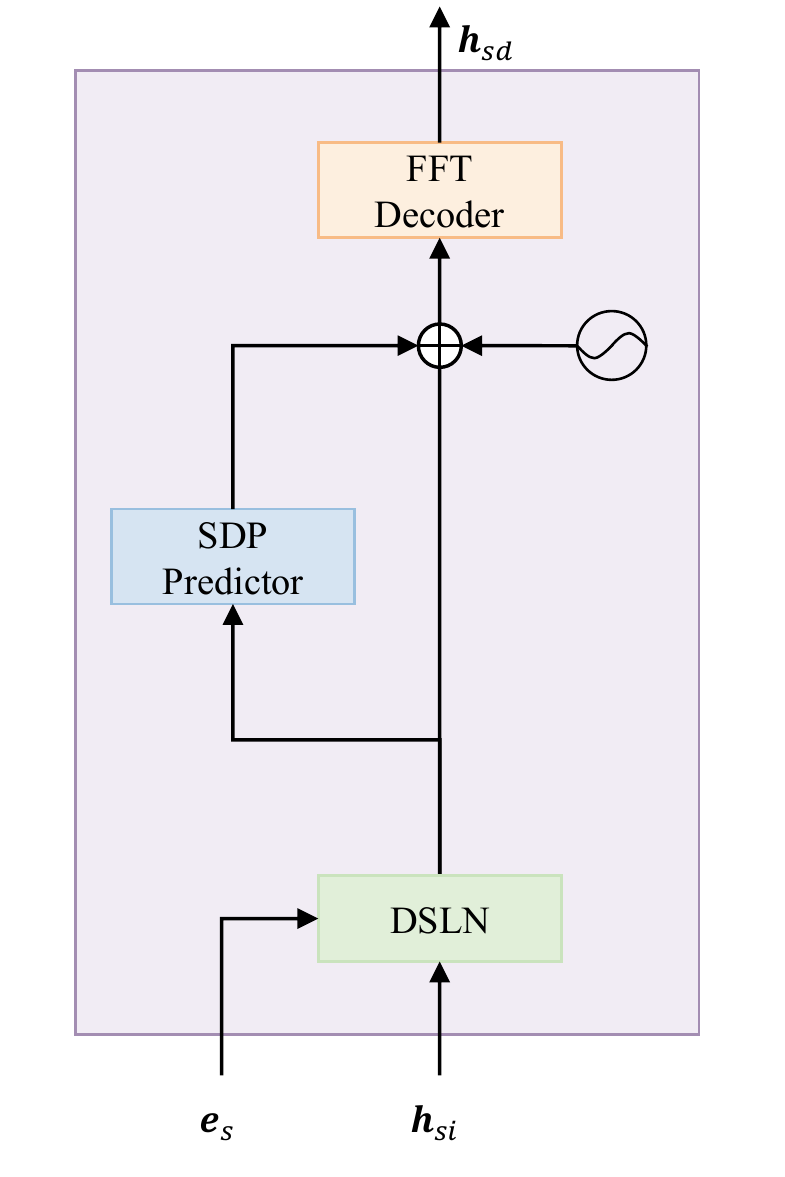}
        \label{fig:ssg}}
    \caption{(a) The overall architecture of CrossSpeech. $\boldsymbol{e}_s$ and $\boldsymbol{e}_l$ refer to speaker and language embedding, respectively. $\boldsymbol{h}_{si}$ and $\boldsymbol{h}_{sd}$ denote the speaker-independent and speaker-dependent acoustic representation, respectively. Also, the sinusoidal symbol means positional encoding. (b) Detailed architecture of the speaker-independent generator. ``M-DSLN" is the mix-dynamic speaker layer normalization and ``SIP" stands for the speaker-independent pitch. (c) Details of the speaker-dependent generator. ``SDP" denotes the speaker-dependent pitch.}
    \label{fig:architecture}
\end{figure*}

CrossSpeech is built upon FastPitch~\cite{lancucki2021fastpitch} with adopting the online aligner~\cite{badlani2022one}. This online aligner not only allows efficient training but also, more importantly, removes the dependency of pre-calculated aligners for each language, which is highly beneficial for extending languages in cross-lingual TTS~\cite{badlani2022one}. Fig.~\ref{fig:architecture} illustrates the overall architecture of CrossSpeech. One intuitive way to avoid speaker-language entanglement in cross-lingual TTS is to divide the generation pipeline into the speaker-independent and speaker-dependent modules~\cite{li2019feature,huang2022generspeech}. Thus, we propose SIG and SDG, which model the speaker-independent and speaker-dependent representations, respectively. Through the mix-dynamic speaker layer normalization (M-DSLN), speaker generalization loss ($\mathcal{L}_{sgr}$) and speaker-independent pitch (SIP) predictor, SIG produces the speaker-independent acoustic representation $\boldsymbol{h}_{si}$ which is not biased to particular speaker distributions. On the top  of SIG, SDG models the speaker-dependent representation $\boldsymbol{h}_{sd}$ through the dynamic speaker layer normalization (DSLN)~\cite{lee2022pvae} and speaker-dependent pitch (SDP) predictor~\cite{lancucki2021fastpitch}. Moreover, we provide $\boldsymbol{h}_{si}$ to the predicted mel-spectrogram via a single projection layer. This residual connection not only improves the pronunciation accuracy but also boosts the speaker and language disentanglement, as validated in~\ref{subsec:ablation}.

\section{Speaker-independent Generator}
In order to produce the generalizable speaker-independent representation, we design SIG including the M-DSLN and speaker generalization loss. Moreover, to learn speaker-independent prosodic variations, we construct the SIP predictor which improves the pitch accuracy and naturalness of the synthesized cross-lingual speech.

\subsection{Speaker Generalization}
\label{sec:spkgen}
Instead of conditioning speaker embedding simply through either summation or concatenation, PVAE-TTS~\cite{lee2022pvae} proposes DSLN which can adaptively modulate hidden features based on the speaker embedding. Given the hidden representation $\boldsymbol{h}$ and speaker embedding $\boldsymbol{e}_s$, the speaker-conditioned representation can be derived as follows:
\begin{equation}
    \text{DSLN}(\boldsymbol{h},\boldsymbol{e}_s) = \mathbf{W}(\boldsymbol{e}_s) \circledast \text{LN}(\boldsymbol{h}) + \mathbf{b}(\boldsymbol{e}_s),
\end{equation}
where $\circledast$ denotes 1d convolution and LN refers to layer normalization. The filter weight $\mathbf{W}(\boldsymbol{e}_s)$ and bias $\mathbf{b}(\boldsymbol{e}_s)$ are predicted by a single linear layer taking $\boldsymbol{e}_s$ as an input.

Inspired by recent works~\cite{huang2022generspeech,zhou2021domain}, we extend DSLN to mix-DSLN (M-DSLN). Through M-DSLN, we can prevent text encoding from being biased to specific speaker attributes, and ensure generalization capability. To this end, we first mix the filter weight and bias as follows: 
\begin{equation}
    \mathbf{W}_{mix}(\boldsymbol{e}_s) = \gamma \mathbf{W}(\boldsymbol{e}_s) + (1-\gamma)\mathbf{W}(\tilde{\boldsymbol{e}}_s),
\end{equation}
\begin{equation}
    \mathbf{b}_{mix}(\boldsymbol{e}_s) = \gamma \mathbf{b}(\boldsymbol{e}_s) + (1-\gamma)\mathbf{b}(\tilde{\boldsymbol{e}}_s),
\end{equation}
where $\tilde{\boldsymbol{e}}_{s}$ is acquired by randomly shuffling $\boldsymbol{e}_s$ along the batch dimension~\cite{huang2022generspeech,zhou2021domain}, and $\gamma$ is sampled from the Beta distribution: $\gamma \sim \text{Beta}(\alpha,\alpha)$. In our experiments,  $\alpha$ is set to $2$. Utilizing the mixed speaker information, we can formulate M-DSLN as follow:
\begin{equation}
    \text{M-DSLN}(\boldsymbol{h}_t,\boldsymbol{e}_s) = \mathbf{W}_{mix}(\boldsymbol{e}_s) \circledast \text{LN}(\boldsymbol{h}_t) + \mathbf{b}_{mix}(\boldsymbol{e}_s),
\end{equation}
where $\boldsymbol{h}_t$ denotes the hidden text representation predicted by the text encoder.

To further increase the generalization performance, we present speaker generalization loss ($\mathcal{L}_{sgr}$) utilizing Kullback-Leibler (KL) divergence. This ensures consistency between the text encoding conditioned on the mixed speaker information and that on the original speaker information~\cite{kang2022style}:
\begin{equation}
    \mathcal{L}_{sgr}^{o2m} = \text{KL}(\text{DSLN}(\boldsymbol{h}_t,\boldsymbol{e}_s) ||\text{M-DSLN}(\boldsymbol{h}_t,\boldsymbol{e}_s)),
\end{equation}
\begin{equation}
    \mathcal{L}_{sgr}^{m2o} = \text{KL}(\text{M-DSLN}(\boldsymbol{h}_t,\boldsymbol{e}_s)||\text{DSLN}(\boldsymbol{h}_t,\boldsymbol{e}_s)).
\end{equation}
Here, we denote $\mathcal{L}_{sgr}=\mathcal{L}_{sgr}^{o2m}+\mathcal{L}_{sgr}^{m2o}$ for brevity. By adopting M-DSLN and speaker generalization loss, we can detach the speaker-dependent information from the linguistic representation, and enable the following SIP and duration predictors to predict speaker-independent variations. 

\subsection{Speaker-independent Pitch Predictor}
\label{sec:sip}
In cross-lingual TTS, it is hard to predict speech variation owing to the speaker-language combinations unseen during training. To alleviate this issue, we introduce the SIP predictor to predict text-related pitch variation which is a common attribute across multiple speakers. Taking the output of M-DSLN as an input, the SIP predictor predicts binary pitch contour sequence which implies a rise or fall of the pitch value~\cite{zhan21_interspeech}.

To train the SIP predictor, we first extract the ground truth pitch value for every frame following FastPitch~\cite{lancucki2021fastpitch}. Here, the ground truth pitch is a speaker-dependent value, and thus we refer to the ground truth pitch sequence as the speaker-dependent pitch sequence $\boldsymbol{p}^{(d)}$. Since the SIP predictor handles pitch values at the input token-level, we average $\boldsymbol{p}^{(d)}$ over every input token using the ground truth duration. At final, we obtain the speaker-independent target pitch sequence $\boldsymbol{p}^{(i)}$ by converting the averaged $\boldsymbol{p}^{(d)}$ into a binary sequence, which can be formulated as follows:
\begin{equation}
    {p}^{(i)}_n = 
    \begin{cases}
        1, & {\bar{p}}^{(d)}_{n-1}<\bar{p}^{(d)}_n \\
        0, & \mbox{otherwise},
    \end{cases}
\end{equation}
where $\bar{p}^{(d)}_n$ means the $n^{th}$ value of averaged $\boldsymbol{p}^{(d)}$ while $p^{(i)}_n$ indicates the $n^{th}$ value of $\boldsymbol{p}^{(i)}$. Here, $n\in\{1,2,3,\cdots,N\}$ where $N$ refers to the length of input tokens. We use binary cross-entropy loss to optimize the SIP predictor:

\begin{equation}
    \mathcal{L}_{sip} = - \sum^{N}_n\big[p^{(i)}_n \text{log}\hat{p}^{(i)}_n + (1-p^{(i)}_n) \text{log}(1-\hat{p}^{(i)}_n)\big],
\end{equation}
where $\hat{p}_{n}^{(i)}$ means the $n^{th}$ predicted speaker-independent pitch value. 

As illustrated in Fig.~\ref{fig:sag}, the speaker-independent pitch sequence is added to the hidden sequence after being passed to the 1d convolutional layer. The resulting sum is upsampled based on token duration and passed to the feed-forward transformer (FFT) decoder~\cite{lancucki2021fastpitch} which converts the upsampled hidden sequence into the speaker-independent acoustic representation $\boldsymbol{h}_{si}$. Note that the duration predictor in CrossSpeech learns \textit{general} duration information because it takes speaker-generalized representation as an input. As proven in the recent study~\cite{cho2022sane}, this leads to predict token duration independent from speaker identity and stabilize the duration prediction in cross-lingual TTS.

\section{Speaker-dependent Generator}
To model speaker-dependent attributes, we construct SDG which consists of DSLN~\cite{lee2022pvae} and SDP~\cite{lancucki2021fastpitch} predictor. DSLN~\cite{lee2022pvae} takes the speaker embedding $\boldsymbol{e}_s$ and speaker-independent acoustic representation $\boldsymbol{h}_{si}$ as inputs, and produces the speaker-adapted hidden feature. Utilizing this speaker-adapted hidden feature, the SDP predictor produces the speaker-dependent pitch embedding at the frame-level.

We extract the speaker-dependent target pitch sequence $\boldsymbol{p}^{(d)}$ as described in Sec.~\ref{sec:sip}, and optimize the SDP predictor using MSE loss:
\begin{equation}
    \mathcal{L}_{sdp} = ||\boldsymbol{p}^{(d)}-\boldsymbol{\hat{p}}^{(d)} ||_2,
\end{equation}
where $\boldsymbol{\hat{p}}^{(d)}$ denotes the predicted speaker-dependent pitch sequence. As illustrated in Fig.~\ref{fig:ssg}, the speaker-dependent pitch sequence is passed to the 1d convolutional layer and summed to the hidden sequence. Finally, The FFT decoder produces speaker-dependent acoustic representation $\boldsymbol{h}_{sd}$ from the adapted hidden sequence. 

To sum up, the overall training objectives ($\mathcal{L}_{tot}$) are given by:
\begin{equation}
    \begin{split}
    \mathcal{L}_{tot} = &\mathcal{L}_{rec} + \mathcal{L}_{align} + \lambda_{dur}\mathcal{L}_{dur} \\ &+ \lambda_{sgr}\mathcal{L}_{sgr}+ \lambda_{sip}\mathcal{L}_{sip} + \lambda_{sdp}\mathcal{L}_{sdp},
    \end{split}
\end{equation}
where $\mathcal{L}_{rec}$ means MSE loss between the target and the predicted mel-spectrogram, $\mathcal{L}_{align}$ denotes the alignment loss for the online aligner described in~\cite{badlani2022one}. Also, $\mathcal{L}_{dur}$ is MSE loss between the target and the predicted duration. In our experiments, we fix $\lambda_{dur}=\lambda_{sgr}=\lambda_{sip}=\lambda_{sdp}=0.1$.

\begin{table}[t]
  \centering
  \caption{Dataset description.}
  \begin{tabular}{cccc}
    \toprule
     Language    & Source   &\#speakers &Hours  \\
    \cmidrule(lr){0-3}
    \multicolumn{1}{c}{\multirow{2}{*}{EN}}  & LJSpeech~\cite{ljspeech17}  &1 &5.98    \\
    \multicolumn{1}{c}{} & Internal &1 &5.59\\
    \cmidrule(lr){0-3}
    ZH &Internal &2 & 11.68\\
    \cmidrule(lr){0-3}
    KO &Internal &12 &9.09 \\
    \bottomrule
  \end{tabular}
  \label{table:data}
\end{table}

\section{Experiments}
We conducted experiments on the mixture of the monolingual dataset in three languages: English (EN), Chinese (ZH), and Korean (KO). The details of dataset are listed in Table~\ref{table:data}. All audios were sampled to $22,050$ Hz and the corresponding transcripts were converted to IPA symbols~\cite{Bernard2021}. 80 bins mel-spectrogram was transformed with a window size of 1024, a hop size of 256, and Fourier transform size of 1024. We used LAMB optimizer~\cite{you2019large} with $\beta_1=0.9$, $\beta_2=0.98$, $\epsilon=1e-9$, a batch size of 16, and followed the same learning rate scheduling method in FastPitch~\cite{lancucki2021fastpitch}.

We mainly followed the detailed model configurations in FastPitch~\cite{lancucki2021fastpitch}. The SIP and SDP predictor followed the pitch prediction pipeline in FastPitch~\cite{lancucki2021fastpitch}, and each FFT decoder was composed of 3 FFT blocks~\cite{vaswani2017attention}. We compared CrossSpeech against recent cross-lingual TTS systems trained on the same configurations: Y. Zhang \textit{et al.}~\cite{zhang2019learning}, D. Xin \textit{et al.}~\cite{xin2021disentangled}, and SANE-TTS~\cite{cho2022sane}. For a fair comparison, the backbone network was unified to FastPitch~\cite{lancucki2021fastpitch}. All the models were trained for 800 epochs, and the predicted mel-spectrogram was converted to audible waveform by pre-trained Fre-GAN~\cite{kim21f_interspeech}. The audio samples are available at our demo page\footnote{\url{https://lism13.github.io/demo/CrossSpeech}}.

\subsection{Quality Comparison}
\begin{table*}[t]
\centering
\caption{Evaluation results. MOS and SMOS are presented with $95\%$ confidence interval. EER denotes speaker verification equal error rate and Cos-sim refers to the x-vector cosine similarity. Lower is better for EER, and higher is better for the other metrics. The bold value represents the best score for each metric.}
\begin{tabular}{lcccccccc}
\toprule
\multirow{2}{*}{\bfseries Method}     & \multicolumn{4}{c}{\bfseries Cross-lingual} & \multicolumn{4}{c}{\bfseries Intra-lingual} \\ \cmidrule(lr){2-5}\cmidrule(lr){6-9}
      & MOS   & SMOS &EER  &Cos-sim & MOS &SMOS & EER  &Cos-sim   \\ \cmidrule(lr){1-9}
Ground truth    &$-$       &$-$        &$-$ &$-$ &$4.39\pm0.05$   & $4.69\pm0.05$ & $3.7\%$  &$0.8519$       \\
Vocoded &$-$       &$-$     &$-$      &$-$ &$4.20\pm0.05$     & $4.60\pm0.05$   & $4.3\%$ & $0.8511$      \\\cmidrule(lr){1-9}
FastPitch &$3.70\pm0.06$        &$2.67\pm0.09$  &$34.6\%$  &$0.6678$       & $3.82\pm0.07$            &$3.83\pm0.08$ &$5.7\%$ & $0.8280$    \\\cmidrule(lr){1-9}
Y. Zhang \textit{et al.} &$3.69\pm0.08$       &$3.61\pm0.08$       &$13.7\%$           &$0.7419$ & $3.80\pm0.07$ & $3.69\pm0.07$ &$6.0\%$&$0.8188$    \\
D. Xin \textit{et al.} &$\mathbf{4.04}\pm\mathbf{0.06}$       &$2.80\pm0.09$       &$42.0\%$           &$0.6180$ &${3.90}\pm{0.08}$ & $3.78\pm{0.07}$ &$\mathbf{5.0}\%$   &$\mathbf{0.8231}$  \\
SANE-TTS&$3.63\pm0.07$       & $3.51\pm0.09$       &$26.7\%$           & $0.7217$ & $\mathbf{3.91}\pm\mathbf{0.08}$ & $3.75\pm0.08$    &$6.3\%$ &$0.8213$ \\ 
CrossSpeech  &$3.72\pm0.08$ &$\mathbf{3.85}\pm\mathbf{0.07}$ &$\mathbf{~~7.0\%}$  &$\mathbf{0.7768}$ & $3.83\pm0.08$ & $\mathbf{3.79}\pm\mathbf{0.08}$ &$5.7\%$ & $0.8197$ \\
\bottomrule
\label{table:compare}
\end{tabular}
\end{table*}

\begin{table}[t]
\centering
\caption{CMOS and CSMOS results of an ablation study. ``Res" refers to the residual connection from $\boldsymbol{h}_{si}$.}
\begin{tabular}{lcccc}
\toprule
\multirow{2}{*}{\bfseries Method}     & \multicolumn{2}{c}{ \bfseries Cross-lingual} & \multicolumn{2}{c}{  \bfseries Intra-lingual} \\ \cmidrule(lr){2-3}\cmidrule(lr){4-5}
      &CMOS   & CSMOS  & CMOS & CSMOS    \\ \cmidrule(lr){1-5}

CrossSpeech &~~~$0.00$       &~~~$0.00$ &~~~$0.00$  &~~~$0.00$     \\ 
    \cmidrule(lr){1-5}
    \emph{w/o} M-DSLN  &$-2.17$  &$-0.56$ & $-0.54$ & $-0.23$\\
    \emph{w/o} $\mathcal{L}_{sgr}$  &$-0.55$ &$-0.80$ &$-0.59$ &$-0.16$\\
    \emph{w/o} SIP &$-0.42$ &$-0.68$ &$-0.61$ &$-0.41$ \\
    \emph{w/o} SDP      &$-0.57$ &$-0.65$ &$-0.55$ &$-0.28$\\  
    \emph{w/o} Res      &$-0.66$ &$-0.73$ &$-0.77$ &$-0.25$ \\
\bottomrule
\label{table:ablation}
\end{tabular}
\end{table}

We assessed CrossSpeech on both subjective and objective evaluations. As a subjective test, we performed 5-scale mean opinion score (MOS) and similarity MOS (SMOS) tests where 30 native speakers were asked to rate the naturalness and speaker similarity on a score of 1-5. For the objective metrics, we computed speaker verification equal error rate (EER) from pre-trained speaker verification network~\cite{jung2022pushing}, and cosine similarity (Cos-sim) between x-vectors~\cite{snyder2018x} extracted from the ground-truth and generated audio. Here, we performed the MOS test to assess speech naturalness, and the other evaluations to measure speaker similarity to the target speaker. We used 50 and 300 samples for subjective and objective tests, respectively.

The results are listed in Table~\ref{table:compare}. Above all, CrossSpeech achieves significant improvements in cross-lingual speech. In cross-lingual TTS, CrossSpeech obtains the best scores in SMOS as well as EER and Cos-sim. This implies CrossSpeech can generate a highly similar voice to the target speaker, outperforming standard methods in terms of speaker similarity. While D. Xin \textit{et al.}~\cite{xin2021disentangled} acquires better naturalness than CrossSpeech in cross-lingual scenarios, it shows poor speaker similarity. We presume it comes from the remaining entanglement between speaker and language distributions, which constrains the model from applying the target speaker identity. Moreover, CrossSpeech shows comparable quality for intra-lingual TTS, verifying that CrossSpeech can enhance the quality of cross-lingual TTS without degrading the quality of intra-lingual TTS.

\subsection{Ablation Study}
\label{subsec:ablation}
To investigate the effect of each CrossSpeech component, we conducted an ablation study using 7-scale comparative MOS (CMOS) and comparative SMOS (CSMOS). In CMOS and CSMOS, judges listened to the audio samples from two systems and compared their naturalness (or similarity) with a score from -3 to 3.  

Table~\ref{table:ablation} shows the results of an ablations study. Here, ``\textit{w/o} Res" refers to ``without residual connection from $\boldsymbol{h}_{si}$." As demonstrated, each component contributes to enhancing the quality of CrossSpeech, both in cross-lingual and intra-lingual scenarios. In particular, the model cannot synthesize intelligible cross-lingual speech after replacing M-DSLN with DSLN. In comparison, it shows a relatively small degradation in intra-lingual speech, which proves the effectiveness of M-DSLN in disentangling speaker and language information in cross-lingual TTS. Removing $\mathcal{L}_{sgr}$ shows the worst similarity score in cross-lingual speech. The absence of SIP or SDP also results in quality degradation. Also, removing the residual connection from $\boldsymbol{h}_{si}$ leads to incorrect pronunciation and quality degradation.

\subsection{Acoustic Feature Space}

\begin{figure}[t]
\centering\includegraphics[width = .95\columnwidth]{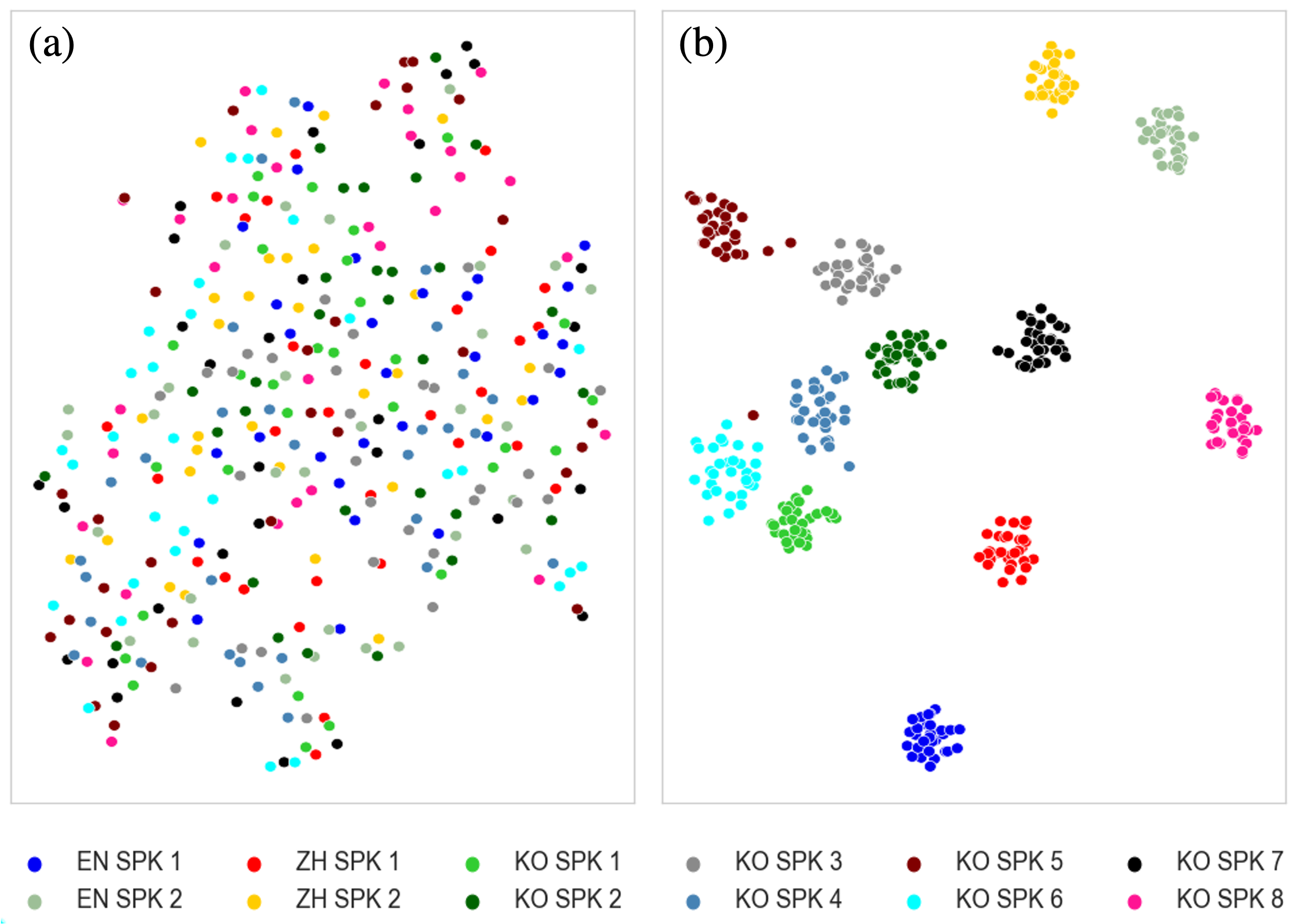}
\caption{t-SNE plots of x-vectors extracted from (a) the projected speaker-independent acoustic representation and (b) the final mel-spectrogram. Each color represents a difference speaker.}
\label{fig:tsne}
\end{figure}

To verify the speaker generalization capability of SIG, we visualized acoustic feature space. Fig.~\ref{fig:tsne} shows t-stochastic neighbor embedding (t-SNE)~\cite{van2008visualizing} plots of x-vectors extracted from (a) the projected speaker-independent acoustic representation and (b) the final mel-spectrogam.

In Fig.~\ref{fig:tsne}(a), we observed that the embeddings are not clustered by speakers but randomly spread out. This indicates that the speaker-independent representations are not biased to speaker-related information but only contain text-related variations. On the other hand, the embeddings are well clustered by speakers in Fig. ~\ref{fig:tsne}(b). This demonstrates that CrossSpeech can successfully learn and transfer speaker-dependent attributes through SDG.

\section{Conclusion}
\label{sec:conclusion}
We proposed CrossSpeech which achieved high-fidelity cross-lingual speech synthesis with significantly improved speaker similarity. We observed insufficient speaker-language disentanglement in previous cross-lingual TTS systems, and addressed the issue by separately modeling speaker and language representations in the level of acoustic feature space. Experimental results demonstrated that CrossSpeech outperformed standard methods while maintaining the quality of intra-lingual TTS. Moreover, we verified the effectiveness of each CrossSpeech component by an ablation study. While there still exists a gap between the target and synthesized audio, we can take advantage of extra features (e.g., energy) to improve the quality of synthesized audio. For future work, we will leverage more external features to improve the generation quality.

\clearpage

\bibliographystyle{IEEEbib}
\bibliography{main}

\end{document}